\documentclass[10pt,conference]{IEEEtran}

\usepackage{cite}
\usepackage{amsmath,amssymb,amsfonts}
\usepackage{booktabs}
\usepackage{enumitem}
\usepackage{graphicx}
\usepackage{url}
\usepackage{xcolor}
\usepackage{listings}
\usepackage{hyperref}
\usepackage{ifthen}
\usepackage{xspace}

\newboolean{commentsenabled}
\setboolean{commentsenabled}{true} 

\ifthenelse{\boolean{commentsenabled}}{
	\newcommand{\nbc}[3]{%
		{\colorbox{#3}{\bfseries\sffamily\scriptsize\textcolor{white}{#1}}}%
		{\textcolor{#3}{\sf\small$\langle$\textit{#2}$\rangle$}}%
	}
	
}{
	\newcommand{\nbc}[3]{}

}

\newcommand{\tool}{\textsc{CC4Q Workbench}\xspace}

\definecolor{codebg}{HTML}{F8FAFC}
\definecolor{codeframe}{HTML}{CBD5E1}
\definecolor{codekw}{HTML}{1E3A8A}
\definecolor{codecmt}{HTML}{64748B}

\begin{document}

\title{\tool: From Quantum Repositories to Analyzable Code Comments}

\author{
\IEEEauthorblockN{Zenghui Zhou, Yuechen Li\textsuperscript{*}, Zheng Zheng}
\IEEEauthorblockA{Beihang University, Beijing, China\\
\{zhouzenghui, liyuechen, zhengz\}@buaa.edu.cn\\
\textsuperscript{*}Corresponding author}
}

\maketitle

\begin{abstract}

Quantum software development kits (SDKs) expose high-level APIs whose use requires an understanding of underlying mathematical principles, circuit semantics, and hardware properties.
Code comments can provide natural-language or structural explanations that are not explicit in API calls.
However, studying comments in a new quantum SDK still requires researchers to coordinate separate steps for comment extraction, quantum-specific classification, and manual verification, resulting in repeated integration effort.
To address this problem, we present \tool, an open-source workbench that integrates these steps into a reusable workflow.
Given a Python repository at a specified revision, \tool automatically extracts code--comment pairs, segments comments into sentence-level units, and applies two trained classifiers to identify quantum-specific content and assign quantum-knowledge categories.
Its web interface supports exploration of bundled datasets, inspection of code snippets, corpus construction from new repositories, manual annotation with model predictions hidden during binary labeling, and classifier fine-tuning using exported annotations.
The release provides aggregate statistics for six SDK datasets comprising 39,593 sentence-level code comment units (SCCUs), of which 17,517 are browsable in the web interface.
Applied to three additional application-layer Qiskit repositories outside the original CC4Q dataset, the pipeline ran without code changes and produced 11,517 SCCUs.
Manual verification of 379 SCCUs showed 85.0\% agreement between human annotations and the binary classifier's predictions.
\end{abstract}

\begin{IEEEkeywords}
Quantum software engineering, Code comments, Software tool
\end{IEEEkeywords}

\section{Introduction}

Quantum SDKs provide high-level APIs that make quantum operations easier to express, but these abstractions do not eliminate the need to understand the concepts and execution semantics behind them. 
In practice, a short API call can hide quantum concepts and implicit behaviors. For example, a call such as
\texttt{qc.cx(0, 1)} does not explain which qubit controls the
operation and which qubit is changed, potentially making it harder for
developers to understand and debug the circuit's behavior. 
Code comments often fill this gap by explaining quantum behavior in natural language next to quantum code~\cite{cc4q2026tosem,yu2025quainth}. 

In this context, an existing study~\cite{cc4q2026tosem} released CC4Q, a dataset of  12,812 extracted code--comment pairs (CCPs) and 28,076 labeled sentence-level code comment units (SCCUs) from three Qiskit repositories, together with comment taxonomies associated with the quantum-specific semantics.
As quantum software ecosystems grow, reusable tooling for code-comment analysis across SDKs is increasingly important. Although the research scripts of~\cite{cc4q2026tosem} are reusable, applying them to another Qiskit repository still requires coordinating separate steps for extracting and linking comments, obtaining labels, and checking their reliability. 
Existing comment-analysis tools for classical software, such as CPC~\cite{zhai2020cse}, target Java API documentation and do not identify quantum-specific knowledge in comments. Researchers therefore lack an integrated workflow for constructing, checking, and adapting quantum comment corpora across repositories.

To fill this gap, we introduce \tool, an open-source workbench for constructing, exploring, and validating quantum code-comment corpora. 
While the companion study~\cite{cc4q2026tosem} focused on comment taxonomies and empirical findings, this paper focuses on the tool itself: how it is built for reuse, and how users work with it through visual interfaces and developer APIs.
This paper makes five contributions:
\begin{itemize}[leftmargin=*]
  \item \textbf{An integrated workbench}: The web interface brings together five components to support corpus exploration, code inspection, repository processing, manual labeling, and model training, respectively.
  \item \textbf{Extensible corpus construction}: The CLI and web workflow process Python repositories by extracting source-linked CCPs and segmenting comments into SCCUs, with support for labeling units using the quantum-relatedness and quantum-category classifiers.
  \item \textbf{Reusable and trainable models}: Users can load released checkpoints for prediction and fine-tune BERT classifiers with built-in or user-provided labeled data.
  \item \textbf{Visual exploration and annotation}: Corpus statistics are visualized to support cross-source comparison, and the annotation interface facilitates manual labeling.
  \item \textbf{Transfer evaluation}: We apply the pipeline without code changes to three previously unstudied application-layer repositories (i.e., \texttt{qiskit-machine-learning}, \texttt{qiskit-nature}, and \texttt{qiskit-algorithms}) yielding 5,502 CCPs and 11,517 SCCUs, while manual checking of 379 SCCUs found 85.0\% agreement between human labels and the binary model's predictions.
\end{itemize}

Paper structure: Section~\ref{sec:architecture} describes \tool{}’s architecture. Section~\ref{sec:usecases} presents three practical use cases, while Section~\ref{sec:validation} evaluates the pipeline’s transferability. Section~\ref{sec:conclusion} concludes with future work.

\section{Tool Architecture}
\label{sec:architecture}
\begin{figure*}[!t]
\centering
\includegraphics[width=0.9\textwidth]{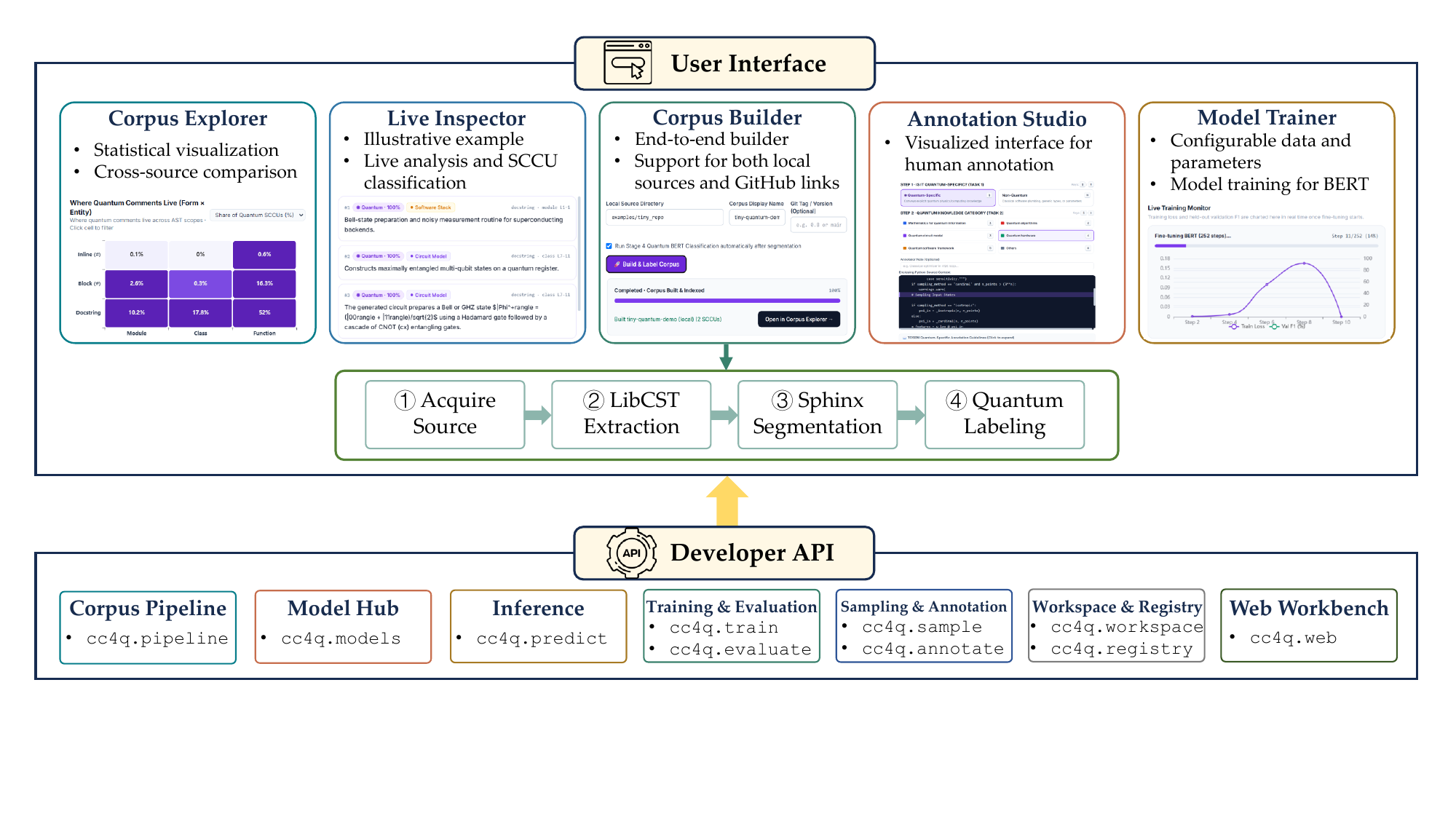}
\vspace{-5em}
\caption{Architecture of \tool.}
\vspace{-1em}
\label{fig:architecture}
\end{figure*}

Fig.~\ref{fig:architecture} summarizes how \tool turns a Python repository
into a labeled corpus of source-linked comments. Users can explore and review
the corpus through the web interface or access the pipeline and related
functions through the developer API.

In the following subsections, we describe the end-to-end builder first (Section~\ref{sec:end-to-end}), then introduce the other four components of the user interface (Section~\ref{sec:ui-components}), and finally present details about the developer API (Section~\ref{sec:dev}).

\subsection{End-to-end Builder}
\label{sec:end-to-end}
\textbf{Corpus Builder}, a core component of \tool{}, provides an end-to-end path from a repository to a
source-linked corpus. It accepts either a local Python directory or a GitHub
repository and can carry the result through comment processing and model
labeling in the same workflow. This removes the manual handoffs between
scripts and intermediate files that otherwise make it difficult to repeat the
analysis for another SDK.

\textbf{Corpus Builder} automatically proceeds through four stages. 
First, users provide a local Python project or a GitHub repository and may specify the version to process. 
The builder then scans the repository’s Python files to extract docstrings, inline comments, and consecutive comment lines, associating each with its source file, line range, and enclosing code entity types (i.e., module, class, or function).
In this stage, LibCST supports this extraction and code-context recovery~\cite{libcst}. Next, the builder converts longer comments into sentence-level units while preserving their source links; docutils~\cite{docutils} and spaCy~\cite{spacy} support the processing of structured docstrings and sentence boundaries. 
Finally, when labeling is enabled, two classifiers perform the two labeling tasks of CC4Q: (1) a binary task labels each unit as quantum-specific or non-quantum, and (2) a six-class task assigns each quantum-specific unit to mathematics for quantum information, quantum algorithms, the quantum circuit model, quantum hardware, quantum software framework, or others. The builder saves the labeled corpus in the workspace, where it can be opened in Corpus Explorer for browsing and comparison.

\par\addvspace{2pt}%
\refstepcounter{lstlisting}\label{lst:record}%
{\centering\footnotesize\lstlistingname~\thelstlisting. Worked example: a class docstring processed by Live Inspector and stored as a labeled corpus record.\par}%
\vspace{2pt}
\begin{lstlisting}[language=Python]
# source: entanglement.py, class EntanglementBuilder (L7-11)
{ "comment_text": "Constructs maximally entangled
    multi-qubit states on a quantum register.",
  "comment_type": "docstring", "code_entity_type": "class",
  "quantum_related": true,
  "quantum_category": "Quantum circuit model" }
\end{lstlisting}

Listing~\ref{lst:record} gives a concrete example of the output. The class
docstring is stored with its comment type, enclosing class, source snippet,
line information, and predicted labels. Keeping these fields together lets a
researcher inspect the code behind a label and move from corpus summaries
back to the original source.





\subsection{Components in User Interface}
\label{sec:ui-components}

The previous subsection described how \textbf{Corpus Builder} turns a repository into a source-linked corpus. 
The other four components support work with that corpus:
researchers can explore its contents, inspect predictions on code snippets,
check labels manually, and adapt a classifier with labeled data. The web
interface is launched with \texttt{cc4q demo} and uses a Starlette/Uvicorn
JSON API~\cite{starlette} and an Apache ECharts frontend~\cite{echarts}.

\textbf{Corpus Explorer} supports browsing and comparing bundled and newly
built corpora. It summarizes quantum-specific shares and category profiles,
and uses heatmaps to show how comments are distributed across comment forms
and code scopes. Selecting a chart value filters the linked comment records,
so users can move from corpus-level patterns to source-linked examples. 

\textbf{Live Inspector} lets users paste Python code or choose a built-in
example for on-demand analysis. After extraction, it displays the comments
split into sentence-level units and their predictions from the currently
active classifiers. For units predicted as quantum-specific, it also shows
the prediction confidence and assigned knowledge category. This provides a
quick way to inspect model behavior on code without building a full corpus.

\textbf{Annotation Studio} supports manual review of model labels. Users can
open the supplied verification sample or upload a \texttt{JSONL} file, then label each
unit as quantum-specific or non-quantum and assign a category to quantum
units. Model predictions are hidden during the binary decision; the predicted
category is offered only as an editable suggestion after a unit is marked
quantum. Labels are saved as users work, and completed sessions can be
exported for evaluation or training.

\textbf{Model Trainer} uses labeled data to adapt either the binary
quantum-relatedness classifier or the six-class category classifier. Users
can select a built-in dataset or upload their own labeled records, then
monitor training loss and validation metrics while training proceeds, where \tool{} locally registers completed models that can be selected for subsequent predictions.

\subsection{Developer API and Command-line Interface}
\label{sec:dev}
Besides the web interface for interactive use, \tool{} offers a developer API so researchers can integrate corpus analysis into automated scripts or extend the tool.
The API exposes the same core capabilities as the workbench, so programmatic use can fit into an existing research workflow.

The Python API provides functions to extract code--comment pairs, segment
comments, and apply the released classifiers to snippets or corpora. It also
supports sampling records for annotation, training classifiers from labeled
data, and evaluating their predictions. Workspace functions persist corpora,
annotation sessions, and evaluation results, while the model registry stores
trained models for reuse.

\par\addvspace{2pt}%
\refstepcounter{lstlisting}\label{lst:cli-commands}%
{\centering\footnotesize\lstlistingname~\thelstlisting. CC4Q command-line entry points.\par}%
\vspace{2pt}
\begin{lstlisting}
# Commands are run separately; arguments are illustrative.
cc4q download-models
cc4q build --repo REPO --output-dir DIR
cc4q predict --input SCCUS --output LABELED
cc4q sample --input DATA --output SAMPLE
cc4q annotate --data SAMPLE --output-dir ANNOTATIONS
cc4q train --task TASK --data LABELS --output-dir MODEL
cc4q evaluate --annotations ANNOTATIONS --key KEY
cc4q demo
cc4q serve --host HOST --port PORT
\end{lstlisting}


The \texttt{cc4q} command-line interface provides a direct way to run these
operations from a terminal; Listing~\ref{lst:cli-commands}
summarizes the available commands. 
A researcher can build a corpus, write predictions to a file, prepare samples for annotation, train or evaluate a classifier, and launch the web workbench with \texttt{cc4q demo}. Explicit input and output paths make these commands
suitable for repeatable batch processing and automated scripts. 

\section{Use Cases}
\label{sec:usecases}
To illustrate how \tool{} supports practical research tasks, we present
three use cases that demonstrate its application to quantum-repository exploration,
codebase analysis, and domain adaptation.

\textbf{Use case 1: Exploring a new quantum repository.}
Beyond the three infrastructure-layer repositories in the original CC4Q dataset, real-world quantum development ecosystems usually include repositories serving specific application purposes.
With \tool{}, for example, a researcher studying the documentation of \texttt{qiskit-nature}, a Qiskit repository designed for quantum chemistry and other natural-science applications, can directly select the repository from the bundled list or provide its GitHub URL to \textbf{Corpus Builder}. The tool can then generate a labeled dataset of 4,842 sentences, with a predicted quantum share of 28.5\%, a comment density of 0.127
(comment lines divided by code lines), and a category profile. The profile
shows that 28.9\% of comments concern \emph{mathematics}, consistent with
the SDK's quantum-chemistry domain. The researcher can export the dataset
in \texttt{JSONL} format for further analysis.

\textbf{Use case 2: Analyzing an existing quantum repository.}
A library maintainer can run the Docker image, point \textbf{Corpus Builder} to a local repository, and inspect the resulting heatmap to examine how quantum knowledge is distributed across code comments. For example, the maintainer
can investigate whether quantum-related knowledge is concentrated in class
docstrings rather than inline comments. The maintainer can also use Live Inspector to analyze a newly developed module and create a labeling session in Annotation Studio to validate the dashboard's predictions on a manually labeled sample.

\textbf{Use case 3: Adapting the classifiers to a new domain.}
We acknowledge that existing classifiers reused from CC4Q may not be generalizable to all quantum repositories, as they were trained on a limited set of infrastructure-layer repositories.
When an SDK combines classical numerical code with quantum primitives, the default classifier may misclassify some comments, potentially degrading data quality. To improve domain adaptation, a research team can manually label a few hundred domain-specific units in Annotation Studio, export the annotations as a \texttt{JSONL} file, and use \textbf{Model Trainer} to fine-tune the classifier from the BERT base model. The adapted model is then
registered locally and becomes available for subsequent corpus builds and inspections.

\section{Transfer Validation}
\label{sec:validation}

The preceding sections described the functionality and use cases of \tool{}, while this section evaluates whether those capabilities
remain dependable in use. 
We therefore address two questions: whether the binary predictions are sufficiently reliable for corpus checking, and whether the complete pipeline transfers to repositories that were not used to develop the models.

\subsection{Using the Classifier on New Qiskit Repositories}

The earlier study evaluated the BERT-based classifiers on test data from three infrastructure-level repositories. 
Here we ask whether these classifiers still work on another Qiskit repository outside the training set.
We then sampled 385 units across the three new repositories (\texttt{qiskit-machine-learning}, \texttt{qiskit-nature}, \texttt{qiskit-algorithms}), 
stratified by SDK, code-entity type, and the model's predicted label. 
Six items were excluded because their model labels had been shown to the annotator during calibration, leaving 379 scored units as the benchmark.
One author then checked the binary labels following the annotation guideline from the earlier study. The model agreed with the human labels on \textbf{85.0\%} of the units. The exact numbers reproduce with:

\begin{lstlisting}
cc4q evaluate \
  --annotations exp/spotcheck_annotation.csv \
  --key exp/spotcheck_answer_key.jsonl \
  --contaminated 1,8,16,25,62,132
\end{lstlisting}

The 57 disagreements contain 32 false positives and 25 false negatives. 
There are two possible reasons: (1) the model sometimes marks classical numerical parts of variational workflows as quantum, such as ODE (ordinary differential equation) solvers and quasi-Newton optimizers used in numerical computations; 
(2) short units that use specialized
abbreviations are misjudged, such as MP2 (second-order M{\o}ller--Plesset perturbation theory), VSCF (vibrational self-consistent field), and FCIDump (a file format for one- and two-electron integrals). 
Not all labels were checked by humans, and the classifiers are not perfect despite their test-set accuracy of 0.936 (binary) and 0.847 (six-class) reported in~\cite{cc4q2026tosem}. Accordingly, \tool{} lets users train or fine-tune the classifiers on their own labeled data.

The web release bundles 361 of these 379 units as an unlabeled labeling demo; the scored benchmark for evaluation remains the full 379-unit set in \texttt{exp/}.

\subsection{Using the Pipeline on New Qiskit Repositories}


\begin{table}[t]
\centering
\caption{Datasets built by \tool{} for three additional repositories, unmodified.
The original three repositories are shown for reference~\cite{cc4q2026tosem}.}
\label{tab:generalization}
\scriptsize
\setlength{\tabcolsep}{2pt}
\begin{tabular}{lclrrrr}
\toprule
Repository & Version & Layer & CCPs & SCCUs & Q.\% & Density \\
\midrule
\texttt{qiskit}     & 2.1.0 & infra. & 9,677 & 21,970 & 35.4 & 0.126 \\
\texttt{qiskit-aer} & 0.17.2 & infra. & 1,153 & 2,368 & 40.6 & 0.114 \\
\texttt{qiskit-ibm-runtime} & 0.45.1 & infra. & 1,982 & 3,738 & 23.5 & 0.210 \\
\midrule
\texttt{qiskit-machine-learning} & 0.9.1 & appl. & 1,663 & 3,140 & 22.4 & 0.172 \\
\texttt{qiskit-nature} & 0.8 & appl. & 2,037 & 4,842 & 28.5 & 0.127 \\
\texttt{qiskit-algorithms} & 0.4.0 & appl. & 1,802 & 3,535 & 25.9 & 0.155 \\
\midrule
\multicolumn{2}{l}{infrastructure total} & &12,812 & 28,076 & \textbf{34.3} & \\
\multicolumn{2}{l}{application total} && 5,502 & 11,517 & \textbf{26.0} & \\
\bottomrule
\end{tabular}
\vspace{-1em}
\end{table}
To assess whether the pipeline transfers without code changes, we ran it on the three repositories listed in Table~\ref{tab:generalization}, where Q.\% is the share of SCCUs that the binary classifier predicts as quantum-specific and Density is the ratio of comment lines to code lines.
Across the six repositories, application-layer repositories show a smaller share of comments
predicted as quantum than infrastructure-layer SDKs (26.0\% vs.\ 34.3\%).
We treat this as a model-labeled observation, not a manually
verified finding. Even so, it suggests that different SDK layers may need
different documentation guidance.

For reproducibility, all SDK release tags, dataset statistics, and validation files are pinned in
the repository. Python dependencies are pinned or lower-bounded in the
packaged requirements. The Docker image and the pytest suite are verified to
build and run in CI. The demo can also start without model checkpoints by
turning off model-dependent actions gracefully.

\section{Conclusion and Future Work}
\label{sec:conclusion}
We present \tool{}, an open-source tool that turns the methodology of CC4Q construction into a reusable workflow. Starting from a Python quantum repository, the tool builds a labeled comment dataset, shows where quantum knowledge appears, lets users check labels without seeing model predictions during labeling, and supports fine-tuning the classifiers on exported labels. 
On three repositories not used in the earlier study, the tool ran without modification, and manual checking found 85.0\% agreement between human labels and the binary model's predictions.

However, \tool{} currently only supports Python-based quantum SDKs such as Qiskit, because stages like automatic comment extraction and segmentation are often language-specific. Future work includes parsers for non-Python SDKs to support more quantum ecosystems.

\section*{Data Availability}
\tool{} is open source under the MIT license at
\url{https://github.com/MaxwelsDonc/cc4q-workbench}, where the repository includes
the Docker image, CLI and Python APIs, SHA256-verified model assets,
six-dataset demo data, the 361-unit labeling demo, and all validation files.
Also, we offer the documentation of \tool{} at \url{https://zhouzenghui.site/cc4q-workbench/}.
The original CC4Q dataset is archived on Figshare~\cite{cc4q2025artifact}.
 

\bibliographystyle{IEEEtran}
\bibliography{ref}

\end{document}